\documentstyle[aps,prl,epsfig]{revtex}
\mathchardef\ogon="012C%
\newcommand{\as}{a\kern-0.22em\lower.40ex\hbox{$_{\ogon}$}}

\begin{document}
\wideabs{
\title{Quantum phases of dipolar bosons in optical lattices}
\author{K. G\'oral$^{1,2}$, L. Santos$^{1}$, and M. Lewenstein$^{1}$}
\address{
(1) Institut f\"ur Theoretische Physik, Universit\"at Hannover, D-30167
Hannover, Germany\\
(2) Center for Theoretical Physics, Polish Academy of Sciences, Aleja 
Lotnik\'ow 32/46, 02-668 Warsaw, Poland}
\maketitle

\begin{abstract}
The ground state of dipolar bosons placed in an optical lattice is 
analyzed. We show that the modification of experimentally accessible 
parameters can lead to the realization and control of different quantum 
phases, including superfluid, supersolid, Mott insulator, checkerboard, 
and collapse phases. 
\end{abstract}

}

The Bose-Einstein condensation (BEC) of dilute atomic gases \cite{bec} has
opened a new interdisciplinary area of modern atomic, molecular and
optical (AMO) physics on one side and 
condensed matter physics on the other: the study of ultracold {\it
weakly interacting} trapped quantum gases \cite{varenna}. 
So far most  of the experiments in this area have been very
accurately described by the semi-classical mean-field method and its
extensions, based on the Gross-Pitaevskii (GP) and Bogoliubov-de
Gennes equations \cite{reviews}. However, experimental 
techniques have recently progressed to a stage at which mean-field methods cease 
to provide an appropriate physical description. In this sense,  
experiments on Feshbach resonances at JILA \cite{JILA} 
allow the modification of the $s$-wave scattering length to such large values that 
the mean-field picture is no more applicable. 
Similarly, the achievement of BEC in metastable helium \cite{He} opens the  
possibility to study higher-order correlation functions, whose analysis 
requires theoretical approaches beyond mean field. 
The recent observation of the Mott insulator-superfluid phase
transition in ultracold atomic samples in optical lattices
\cite{Greiner}, predicted in \cite{Jaksch}, belongs to the same category,
but at the same time initiates a new research area of AMO physics: the physics of
{\it strongly correlated} quantum gases. 
The experiments of \cite{Greiner} are relatively easy 
to accurately control and manipulate and thus 
provide a novel and particularly promising test ground 
for theories of quantum phase transitions 
\cite{Sachdev}, which have traditionally dealt with
condensed-matter systems rather than with atomic gases.

The influence of dipole-dipole forces on the properties of 
BEC has also drawn a considerable attention recently. 
It has been shown that these forces significantly modify 
the ground state and collective excitations 
of trapped condensates \cite{You,Goral,Santos}. 
Dipole-dipole interactions are also responsible for 
spontaneous polarization and spin waves in spinor condensates in optical 
lattices \cite{Meystre} and may lead to self-bound 
structures in the field of a traveling wave \cite{Giovanazzi}. 
In addition, since dipole-dipole interactions can be quite strong 
relative to the short-range (contact) interactions,     
dipolar particles are considered to be promising candidates for the implementation 
of fast and robust quantum-computing schemes \cite{DDgates,qcomp}. 
Sources of cold dipolar bosons include atoms 
\cite{dipolar_atoms} or molecules \cite{dipolar_molecules} with 
permanent magnetic or electric dipole moments. 
Other possible candidates could be atoms 
with electric dipoles, induced either by large 
dc electric fields \cite{You} or by optically admixing 
the permanent dipole moment of a low-lying Rydberg state to the atomic 
ground state in the presence of a moderate dc electric field \cite{Santos,DDgates}.

This Letter is devoted to the analysis of the ground state of an   
ultracold gas of polarized dipolar bosons in an optical lattice. 
The ground state of a gas of short-range repulsively-interacting bosons in a periodic 
potential 
can be either in a superfluid phase or in a Mott-insulating phase, 
characterized by integer boson densities 
and the existence of a gap for particle-hole excitations \cite{Fisher}. 
The superfluid-Mott insulator transition
in cold bosonic atoms in optical lattices has been recently theoretically analyzed 
\cite{Jaksch} and experimentally demonstrated \cite{Greiner}. 
For the case of finite-range interactions new quantum phases have been predicted 
\cite{supersolid}, including supersolid phases which combine both diagonal and off-diagonal 
long-range ordering.  To the best of our knowledge, dipole-dipole interactions 
have not yet been discussed in this context. We show in the following that
these interactions, which are long-range and anisotropic, lead to new interesting 
properties. The long-range character of the dipole-dipole potential provides  
a rich variety of quantum phases. Moreover, we show that 
the interactions in a gas of dipolar bosons are easily tunable, 
allowing for the experimental engineering of quantum phase transitions between 
various kinds of ground states. Such a highly controllable system may be crucial in 
answering some 
unresolved questions in the theory of quantum phase transitions (e.g. the
existence of a  yet-unobserved supersolid \cite{Leggett}, or a Bose metal
at zero temperature \cite{metal}).

A dilute gas of bosons in a periodic potential (e.g. in an optical 
lattice) can be described with the 
help of the Bose-Hubbard (BH) model \cite{Jaksch}. For particles interacting  
via long-range forces the BH Hamiltonian becomes:

\begin{eqnarray} \label{H}
H&=&J\sum_{<i,j>}b_{i}^{\dagger}b_{j}
+\frac{1}{2}U_{0}\sum_{i}n_{i}(n_{i}-1) \nonumber \\
&+&\frac{1}{2}U_{\sigma_1}\sum_{<i,j>}n_{i}n_{j}+\frac{1}{2}U_{\sigma_2}\sum_{<<i,j>>}n_{i}n_{j}
+\ldots,
\end{eqnarray}

\noindent where $b_{i}$ is the annihilation operator of a particle at the  
lattice site $i$, which is considered as being in a state described by 
the Wannier function $w({\bf r}-{\bf r}_{i})$ of the lowest energy band, 
localized on this site. This implies the assumption that the energies 
involved in the system are small compared to the excitation energies 
to the second band. We denote the position of  the local minimum of the optical 
potential as ${\bf r}_{i}$, and the number operator for the site $i$ as 
$n_{i}=b_{i}^{\dagger}b_{i}$. 
In Eq. (\ref{H}) only the nearest-neighbor tunneling  is considered, which
is described by

\begin{equation}
J=\int w^{\star}({\bf r}-{\bf r}_{i}) [-\frac{\hbar^2}{2m}\nabla^{2}+V_{l} 
({\bf r})] w({\bf r}-{\bf r}_{j}) \,{\rm d^3} r \; , 
\end{equation}

\noindent where $j$ and $i$ are indices of the neighboring sites, and $V_{l}({\bf 
r})=\sum_{\xi=x,y,z}V_{\xi}^{0}\cos^{2}(k_{\xi}\xi)$ 
is the optical lattice potential with the wavevector ${\bf k}$. The 
interparticle interactions are characterized by the parameters 

\begin{equation}
U_{\sigma}=\int |w({\bf r}-{\bf r}_{i})|^{2} 
V_{int}({\bf r}-{\bf r'})|w({\bf r'}-{\bf r}_{j})|^{2} \,{\rm d^3} r \; 
{\rm d^3} r' \; ,
\end{equation}

\noindent where $|{\bf r}_i-{\bf r}_j|=4\pi \sigma/|{\bf k}|$. 
$U_0$  determines the on-site interactions, 
$U_{\sigma_1}$ the nearest-neighbor interactions, $U_{\sigma_2}$ 
the interactions between 
the next-nearest neighbors, etc. Consequently, the respective summations
in  Eq.\ (\ref{H}) must be carried out over appropriate pairs of sites
which  are marked by $<\dots >$ for the nearest neighbors, $<<\dots >>$ for the 
next-nearest neighbors, etc. 
In the 2D calculations presented below, we have taken into 
account interactions with up to $4$ neighbors 
($\sigma_1=1$, $\sigma_2=\sqrt{2}$, $\sigma_3=2$, $\sigma_4=\sqrt{5}$), 
since in the particular cases we analyzed the effects of interactions of a longer
range are negligible. In the case of polarized dipoles the interaction
potential is

\begin{equation} \label{inter}
V_{int}=d^{2}\frac{1-3\cos^2\theta}{|{\bf r}-{\bf 
r'}|^3}+\frac{4\pi \hbar^{2}a}{m}\delta({\bf r}-{\bf r'}) \; .
\end{equation}

\noindent where the first term is the dipole-dipole interaction  
characterized by the dipole $d$ and the angle
$\theta$ between the dipole direction and the vector 
${\bf r}-{\bf r'}$, and the second term is the 
short-range interaction given by the $s$-wave scattering length 
$a$  and the atomic mass $m$. 

We find the ground state of the system using a variational approach 
(see \cite{Jaksch} and references therein) based on the Gutzwiller 
ansatz $|\Psi_{MF}\rangle=\prod_{i}|\phi_{i}\rangle$ for the ground-state
wavefunction, where the product is over all lattice sites. The
wavefunctions $|\phi_{i}\rangle$ for each site are expressed in the 
basis of Fock states, $|\phi_{i}\rangle=\sum_{n=0}^{\infty}f_{n}^{i}|n\rangle_{i}$, 
where $n$ indicates the occupation number.  
The coefficients $\{f_{n}^{i}\}$ are found by minimizing 
the expectation value of the Hamiltonian (\ref{H}) in 
the state $|\Psi_{MF}\rangle$ under the constraint of 
a fixed chemical potential $\mu$.

In the following we consider 1D and 2D geometries. 
Low-dimensional BECs have been achieved in recent experiments \cite{Gorlitz} by 
transversally confining a condensate in a tight optical or magnetic 
harmonic trap. A 1D or 2D lattice is created by a laser standing wave, which generates 
a periodic optical potential \cite{Orzel}. 
We have carried out the minimization of 
$\langle \Psi_{MF}|H-\mu\sum_{i}n_{i}|\Psi_{MF}\rangle$ for 
1D lattices with up to $20$ sites and for square 2D lattices with up 
to $9\times 9$ sites. A similar qualitative picture is expected for a 
larger number of lattice sites. Since in 1D and for systems with few atoms the 
application of a mean-field calculation could be questionable (due to the possibly 
important role of fluctuations), we restrict our discussion of the BH 
Hamiltonian (\ref{H}) to the 2D case.

For a square 2D lattice in the $xy$ plane, 
the wavefunctions $|\phi_{i}\rangle$ can be written 
as a product of Wannier functions in the $x$ and $y$ directions and Gaussian 
functions in the $z$ direction. 
There are two generic situations for dipoles in 
2D lattices, namely (i) the dipole is along the $z$ direction or (ii) the 
dipole direction is in the $xy$ plane. This follows from the fact 
that two dipoles experience maximal attraction along the dipole direction and 
maximal repulsion in the transversal plane. As shown in Ref.\ 
\cite{Santos}, the 
mean-field dipole-dipole energy  critically depends on the shape of the 
bosonic cloud, 
which can be altered by modifying the trap. It is intuitively clear that a 
cloud elongated in the 
dipole direction is unstable due to the predominance of attractive 
interactions. On the contrary, the cloud may be stable if it is broader 
in the transversal plane than in the dipole direction. 
In particular, for spherically symmetric wavefunctions $|\phi_{i}\rangle$  the 
on-site averaged dipole-dipole potential vanishes, and only the short-range 
interactions contribute to $U_0$. 

\begin{figure}[ht]
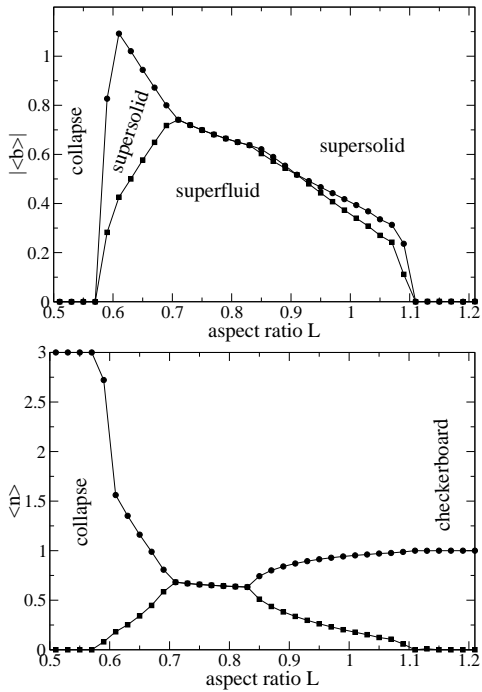
 
\begin{center}
\epsfxsize=5.3cm 
\psfig{file=super.eps,width=6.3cm,clip}
\psfig{file=np.eps,width=6.3cm,clip}
\end{center} 
\caption{Maximal (circles) and minimal (squares) values of the superfluid 
parameter $\langle b_{i} \rangle$ and of the occupation number $\langle 
n_{i} \rangle$ as a function of the aspect ratio $L$ of the on-site 
wavefunctions.
}
\label{aspect}  
\end{figure}

Let us first focus on the case (i). Since the dipole-dipole mean-field 
critically depends on the shape of the cloud, 
the balance between attractive and repulsive interactions 
can be easily manipulated either by modifying the wavelength and intensity of
the lattice or by changing the transversal trapping.
In the following we employ the latter possibility to provide an example of
how different phases of the BH Hamiltonian (\ref{H}) may be accomplished 
just by changing the magnitude of controllable external fields. 
The expectation values $\langle b_{i} \rangle$ provide the superfluid order 
parameter. It is non zero and constant    
for all lattice sites in the superfluid phase, whereas it is periodically 
modulated in the supersolid phase.
Fig. \ref{aspect} shows the maximal (circles) and minimal (squares) value 
of 
$|\langle b_{i} \rangle|$ and of the occupation number $\langle 
n_{i} \rangle$  as a function of the aspect ratio of the 
on-site wavefunction $|\phi_{i}\rangle$. The aspect ratio is defined as 
the square of the ratio between the width of $|\phi_{i}\rangle$ 
in the $x$ direction and the width in the $z$ direction, $L=(l_x/l_z)^2$. 
In the calculations presented in Fig. \ref{aspect}, 
we have considered the case of $\, ^{23}$Na atoms with an 
induced dipole moment of $0.334$ Debye, placed in a lattice of wavelength 
$\lambda=795$ nm. The maximum of the lattice potential is $V_{x,y}^{0}=10 
E_r$, where $E_{r}=\hbar^{2}k^{2}/2m$ is the recoil energy. 
In our simulations we fix the chemical potential $\mu=0.082 \, E_r$, 
which determines 
the mean number of atoms and mean density. 
The quantum phases appearing for the different aspect ratios depend, of course, on the 
chosen physical parameters. However, we have observed 
a similar picture of tunable quantum phases 
for every set of parameters that we have considered.

For $L\ge 1.1$ a checkerboard insulating phase is achieved, in which a 
site occupied by exactly one atom is followed by an empty site. This phase 
is a result of the long-range repulsion between particles in the presence 
of a relatively weak tunneling, which prevents the appearance of a 
superfluid. For $0.9<L<1.1$ the system enters a supersolid phase, 
possessing both diagonal and off-diagonal long-range order, in which 
the system is superfluid, but the superfluid parameter shows a slight 
periodic modulation. For $0.7<L<0.9$ the influence of tunneling 
relative to the long-range interactions is large enough to enforce a 
homogeneous superfluid phase. For $0.57<L<0.7$ the system is a supersolid 
with a strongly modulated superfluid parameter. This phase appears due to 
a significant mutual cancellation of the on-site dipole-dipole 
interactions (attractive for $L<1$) and the always repulsive short-range 
potential -- the system enters an interesting purely long-range regime 
with the local interactions essentially absent. In such a case the 
ratio $J/U_{0}$, which governs the 
insulator-superfluid crossover \cite{Fisher}, increases,  driving the 
system from an insulating phase to a superfluid one. On the other hand,  
the long-range interactions, characterized by the coefficients 
$U_{\sigma_i}$, remain considerably large and
positive. As a consequence, a periodic
modulation of the superfluid parameter occurs.  
Finally, for $L\lesssim 0.57$ the system undergoes local collapses due to 
the attractive local interactions. This last regime has been confirmed by 
checking that the maximal possible occupation per site is always achieved, 
independently of the value of such maximal occupation.

\begin{figure}[ht]
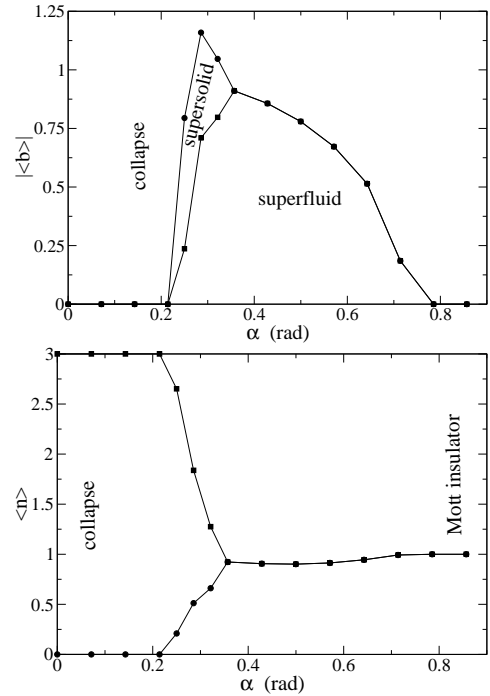
 
\begin{center} 
\epsfxsize=5.3cm
\psfig{file=tilt-super.eps,width=6.3cm,clip}
\psfig{file=tilt-np.eps,width=6.3cm,clip}
\end{center} 
\caption{Maximal (squares) and minimal (circles) values of the superfluid 
parameter $\langle b_{i} \rangle$ and of the occupation number $\langle 
n_{i} \rangle$ as a function of the tilt angle $\alpha$.}
\label{tilt}
\end{figure}

Another simple experimental control knob is provided by the angle $\alpha$ 
between the dipole direction and the vector normal to the 2D lattice 
plane. For $\alpha=0$ we recovered the previous results. For $\alpha>0$ 
the coefficients $U_\sigma$ depend 
not only on the distance between neighbors but also on the angle between 
the projection of the dipole direction on the lattice plane and the vector 
joining the corresponding sites. In Fig. \ref{tilt} we present a sequence 
of quantum phases obtained when the angle $\alpha$ is varied. 
For this calculation, the aspect ratio is fixed to $L=0.5$ and the rest of the 
parameters is kept the same as those of Fig. \ref{aspect}. For $\alpha$ 
approaching $\pi/2$, we observe only Mott insulator phase or superfluid one, 
since when the projection of the dipole onto the lattice plane is
sufficiently large the dipole-dipole on-site interaction 
becomes positive and reinforces the repulsive on-site contact interactions. 
In other words, tilting the dipoles towards the 
lattice plane brings the system back to a situation of dominant 
local interactions \cite{Jaksch,Fisher}. For $\alpha$ approaching zero, we 
observe collapse in this particular case, as expected for $L=0.5$ 
from Fig.\ \ref{aspect}. Additionally, we should point out 
that we do not observe anisotropic  phases, which one might expect due 
to the anisotropy of interactions, since we employ 
periodic boundary conditions in our simulations.

Finally, let us stress that 
in the analysis of the BH Hamiltonian (\ref{H}) we have expanded 
the field operator in the basis of Wannier functions, which are exact 
solutions of the single-particle problem in a periodic potential. 
This method should give correct results as long as the mean occupation 
of sites 
is of the order of unity. However, nowadays it is possible to load large 
Bose-Einstein condensates 
into optical lattices \cite{Orzel}, 
which results in a very high occupation of sites. 
In such a situation, as long as the lattice potential 
does not prevent the establishment of a common phase between sites, the 
GP equation, routinely used to describe the condensate 
wavefunction in harmonic traps, should provide a correct description of 
the system \cite{Santos,Choi}. In the presence of sufficiently 
strong dipole-dipole forces we can neglect the short-range interactions
and the time-independent GP equation reads:
\begin{eqnarray}
\mu \psi({\bf r}) &=& \{-\frac{\hbar^2}{2m}\nabla^2+V_{l}({\bf 
r})+V_{t}({\bf r}) \nonumber \\
&+& d^2\int d{\bf r'}\frac{1-3\cos^2\theta}{|{\bf r}-{\bf r'}|^3}
|\psi({\bf r'})|^2 \} \psi({\bf r}) \; ,
\label{GPE}
\end{eqnarray}
\noindent where $\psi({\bf r})$ is the wavefunction of the condensate 
(normalized to the total number of particles $N$) and 
$V_{t}({\bf r})=m \omega^2 r^2/2$ is a spherically symmetric 
harmonic trap with frequency $\omega$. In the absence of the 
lattice potential $V_{l}({\bf r})$, the condensate is stable as long 
as $\sigma=N\frac{m}{\hbar^2}\sqrt{\frac{m\omega}{\hbar}}d^2$ does not 
exceed some critical value $\sigma_{cr}$ (for a spherical trap 
$\sigma_{cr}=\sigma_{cr}^{0}\approx 4.3$ \cite{Santos}). We have observed 
that, by raising various 1D and 2D lattice configurations, one can either 
destabilize the condensate for $\sigma<\sigma_{cr}^{0}$ 
 or make it stable in the regime $\sigma>\sigma_{cr}^{0}$. 
For instance, the BEC is stabilized for a 1D lattice whose wavevector 
is along the dipole direction, or a 2D lattice on a plane which contains 
the dipole direction. The discussion of these results will be presented in 
detail elsewhere. Let us just stress here that dipolar gases provide also 
in this situation a unique and very efficient possibility of coherent 
control of a BEC.

In this Letter we have analyzed the ground state of dipolar bosons placed in 
an optical lattice. We have shown that by modifying well-controllable 
parameters, different quantum phases can be accomplished, including 
superfluid, supersolid, Mott insulator, checkerboard and collapse phases. 

We acknowledge support from the Alexander von Humboldt Stiftung, the 
Deutscher Akademischer Austauschdienst (DAAD), the Deutsche 
Forschungsgemeinschaft, the RTN Cold Quantum gases, 
ESF Program BEC2000+, and the subsidy of the 
Foundation for Polish Science. We thank B. Altschuler, M. Baranov, I. 
Bloch, {\L}. Dobrek, Z. Idziaszek, K. Rz{\as}\.zewski, G. Sch\"on, and 
P. Zoller for discussions. K.G. is grateful to D. DeMille for an initial 
stimulus to this work. Part of the results was obtained using 
computers at the Interdisciplinary Center for Mathematical and 
Computational Modeling at Warsaw University.


\begin{references}

\bibitem{bec} M. J. Anderson {\it et al}., Science {\bf 269},
198 (1995);  K. B. Davis {\it et al}., Phys. Rev. Lett. {\bf 75}, 3969
(1995); C.C. Bradley {\it et al.}, {\it ibid.} {\bf 75}, 1687 (1995); {\bf 
79}, 1170(E) (1997).

\bibitem{varenna}
{\it Bose-Einstein Condensation in Atomic Gases}, Proceedings of the
International School of Physics "Enrico Fermi", Course CXL, edited by
M. Inguscio, S. Stringari, C.E. Wieman (IOS Press, Amsterdam, 1999).

\bibitem{reviews}
F. Dalfovo {\it et al}.,
Rev. Mod.  Phys. {\bf 71}, 463 (1999).

\bibitem{JILA} E. A. Donley {\it et al.}, Nature (London) {\bf 412}, 295 
(2001).

\bibitem{He}
A. Robert {\it et al}.,
Science {\bf 292}, 461 (2001);
F. Pereira Dos Santos {\it et al}.,
Phys. Rev. Lett. {\bf 86}, 3459 (2001).

\bibitem{Greiner}
M. Greiner {\it et al.}, Nature (London), {\bf 415}, 39 (2002).

\bibitem{Jaksch}
D. Jaksch {\it et al.}, Phys. Rev. Lett. {\bf 81}, 3108 (1998).

\bibitem{Sachdev}
S. Sachdev, {\it Quantum Phase Transitions} (Cambridge University Press, 
New York, 2000).

\bibitem{You} S. Yi and L. You, Phys. Rev. A {\bf 61}, 041604 (2000); 
S. Yi and L. You, {\it ibid.} {\bf 63}, 053607 (2001); 
S. Yi and L. You, cond-mat/0111256.

\bibitem{Goral} 
 K. G\'oral, K. 
Rz{\as}\.zewski, and T. Pfau, {\it ibid.} {\bf 61}, 051601 (2000);  J.-P. 
Martikainen, M. Mackie, and K.-A. Suominen, {\it ibid.} {\bf 64}, 037601 
(2001).

\bibitem{Santos}
L. Santos {\it et al}.,
Phys. Rev. Lett. {\bf 85}, 1791 (2000).

\bibitem{Meystre}
H. Pu, W. Zhang, and P. Meystre, Phys. Rev. Lett. {\bf 87}, 140405 (2001); 
W. Zhang {\it et al}., {\it ibid.} {\bf 88}, 060401 (2002).

\bibitem{Giovanazzi}
S. Giovanazzi, D.O'Dell, and G. Kurizki, Phys. Rev. Lett. {\bf 88}, 130402 
(2002).

\bibitem{DDgates}
D. Jaksch {\it et al.}, Phys. Rev. Lett. {\bf 85}, 2208 (2000).

\bibitem{qcomp}
G.K. Brennen {\it et al.}, Phys. Rev. Lett. {\bf 82}, 1060 (1999); G.K. 
Brennen, I.H. Deutsch, and C.J. Williams, Phys. Rev. A {\bf 65}, 022313 
(2002); D. DeMille, Phys. Rev. Lett. {\bf 88}, 067901 (2002).

\bibitem{dipolar_atoms}
J. D. Weinstein {\it et al}.,
Phys. Rev. A {\bf 57}, R3173 (1998); 
J. Stuhler {\it et  al.}, {\it ibid.} {\bf 64}, 031405 (2001).

\bibitem{dipolar_molecules}
H.L. Bethlem {\it et al}.,
Nature (London) {\bf 406}, 491 (2000).

\bibitem{Fisher}
M. P. A. Fisher {\it et al.}, Phys. Rev. B {\bf 40}, 546 (1989).

\bibitem{supersolid} 
C. Bruder {\it et al.}, Phys. Rev. B {\bf 47}, 342 (1993); 
A. van Otterlo and K.-H. Wagenblast, Phys. Rev. Lett. {\bf 72}, 3598 (1994); 
G.G. Batrouni {\it et al.},  Phys. Rev. Lett. {\bf 74}, 2527 (1995);
T. D. K\"uhner, S. R. White, and H. Monien, Phys. Rev. B {\bf 61}, 12474 (2000).


\bibitem{Leggett}
A.F. Andreev and I.M. Lifshitz, Sov. Phys. JETP {\bf 29}, 1107 (1969);
G.V. Chester, Phys. Rev. A {\bf 2}, 256 (1970);
A.J. Leggett, Phys. Rev. Lett. {\bf 25}, 1543 (1970).

\bibitem{metal}
D. Das and S. Doniach, Phys. Rev. B {\bf 60}, 1261 (1999); D. Dalidovich 
and P. Phillips, {\it ibid.} {\bf 64}, 052507 (2001).

\bibitem{Gorlitz}
A. G\"{o}rlitz {\it et al.}, Phys. Rev. Lett. {\bf 87}, 130402 (2001);
F. Schreck {\it et al.}, {\it ibid.} {\bf 87}, 080403 (2001);
H. Ott {\it et al.}, {\it ibid.} {\bf 87}, 230401 (2001);
W. H\"ansel {\it et al.}, Nature (London) {\bf 413}, 498 (2001);
S. Burger {\it et al.}, Europhys. Lett. {57}, 1 (2002).

\bibitem{Orzel} 
B. P. Anderson and M. A. Kasevich, Science {\bf 282}, 1686 (1998);
C. Orzel {\it et al.}, {\it ibid.} {\bf 291}, 2386 (2001);
F. S. Cataliotti {\it et al.}, {\it ibid.} {\bf 293}, 843 (2001);
S. Burger {\it et al.}, Phys. Rev. Lett. {\bf 86}, 4447 (2001);
O. Morsch {\it et al.}, {\it ibid.} {\bf 87}, 140402 (2001);
M. Greiner {\it et al.}, {\it ibid.} {\bf 87}, 160405 (2001).

\bibitem{Choi}
D.-I. Choi, and Q. Niu, Phys. Rev. Lett. {\bf 82}, 2022 (1999).

\end{references}
\end{document}